\documentclass{llncs}

\usepackage{t1enc} 
\usepackage{amsmath}
\usepackage{latexsym}
\usepackage{theorem}
\usepackage{amssymb}
\usepackage{url}
\usepackage{wasysym} 

\newcommand{\comment}[1]{}

\newcommand{\vsp}[1][3mm]{\vspace*{#1}}
\newcommand{\moins}{\setminus}
\newcommand{\vide}{\emptyset}
\newcommand{\ie}{{\em i.e.} }
\newcommand{\eg}{{\em e.g.} }

\newcommand{\dom}{\mr{dom}}
\newcommand{\codom}{\mr{codom}}
\newcommand{\FV}{\mr{FV}}
\newcommand{\pos}{\mr{Pos}}

\renewcommand{\a}{\rightarrow}
\newcommand{\A}{\Rightarrow}

\newcommand{\ad}{\downarrow}

\newcommand{\au}{\uparrow}
\renewcommand{\to}{\mapsto}
\newcommand{\al}{\leftarrow}

\newcommand{\I}[1]{[\![#1]\!]}

\newcommand{\ex}{\exists}
\newcommand{\all}{\forall}
\newcommand{\ou}{\vee}

\newcommand{\et}{\wedge}

\newcommand{\sle}{\subseteq}
\newcommand{\sge}{\supseteq}

\newcommand{\lex}{_\mr{lex}}
\newcommand{\mul}{_\mr{mul}}

\renewcommand{\b}{\beta}

\renewcommand{\d}{\delta}

\newcommand{\vep}{\varepsilon}

\renewcommand{\t}{\theta}

\renewcommand{\l}{\lambda}

\newcommand{\s}{\sigma}
\renewcommand{\S}{\Sigma}

\newcommand{\w}{\omega}

\newcommand{\mc}{\mathcal}

\newcommand{\mr}{\mathrm}
\newcommand{\mb}{\mathbb}

\newcommand{\mf}{\mathfrak}
\newcommand{\ms}{\mathsf}

\newcommand{\bN}{\mb{N}}

\newcommand{\bT}{\mb{T}}

\newcommand{\cB}{\mc{B}}

\newcommand{\cF}{\mc{F}}

\newcommand{\cI}{\mc{I}}

\newcommand{\cQ}{\mc{Q}}

\newcommand{\cT}{\mc{T}}

\newcommand{\cX}{\mc{X}}

\newcommand{\ka}{\mf{a}} 

\newcommand{\fB}{\ms{B}}

\newcommand{\fD}{\ms{D}}

\newcommand{\fN}{\ms{N}}

\newcommand{\fP}{\ms{P}}

\newcommand{\va}{{\vec{a}}}
\newcommand{\vb}{{\vec{b}}}

\newcommand{\vl}{{\vec{l}}}

\newcommand{\vr}{{\vec{r}}}

\newcommand{\vt}{{\vec{t}}}
\newcommand{\vu}{{\vec{u}}}
\newcommand{\vv}{{\vec{v}}}

\newcommand{\vA}{{\vec{A}}}
\newcommand{\vB}{{\vec{B}}}

\newcommand{\vT}{{\vec{T}}}
\newcommand{\vU}{{\vec{U}}}

\newenvironment{rul}
  {$\begin{array}{rcl}}
  {\end{array}$}

\newenvironment{rew}[1][~~\a~~]
  {$\begin{array}{r@{#1}l}}
  {\end{array}$}
\newenvironment{rewc}[1][~~\a~~]
  {\begin{center}\begin{rew}[#1]}
  {\end{rew}\end{center}}

  {\end{array}$\end{center}}

{\theorembodyfont{\rmfamily} 

}

\newenvironment{lstgeneric}[2]
  {\begin{list}{#1}{\topsep=.5mm\itemsep=.5mm\parsep=0mm%
    \itemindent=-3ex\labelsep=1ex\labelwidth=0ex #2}}
  {\end{list}}

\newenvironment{lst}[1]
  {\begin{lstgeneric}{#1}{\itemindent=-1ex}}
  {\end{lstgeneric}}

\newenvironment{enumi}[1]
  {\begin{lstgeneric}{}{\usecounter{enumi}\leftmargin=7mm%
    }}
  {\end{lstgeneric}}

\newcommand{\SN}{\mr{SN}}
\newcommand{\CC}{\mr{CC}}
\newcommand{\CR}{\mr{CR}}
\newcommand{\Acc}{\mr{Acc}}

\renewcommand{\prod}{_\mr{prod}}

\newcommand{\horpo}{_{\mr{horpo}}}
\newcommand{\horco}{_{\mr{horco}}}
\newcommand{\whorco}{_{\mr{whorco}}}
\newcommand{\stat}[1]{_{\mr{stat}_{#1}}}

\newcommand{\red}[1]{{\a\!\!(#1)}}

\newcommand{\ab}{\a_\b}
\newcommand{\ar}{\a_R}
\renewcommand{\ae}{\a_E}
\newcommand{\are}{\a_{R,E}}
\newcommand{\arbe}{\a_{R,\b\eta}}
\newcommand{\eqbe}{=_{\b\eta}}
\newcommand{\abz}{\a_{\b_0}}

\newcommand{\lx}{\l x}
\newcommand{\ly}{\l y}

\begin{document}


\title{Computability Closure: Ten Years Later}

\author{Fr\'ed\'eric Blanqui}

\institute{INRIA\\
LORIA\thanks{UMR 7503 CNRS-INPL-INRIA-Nancy2-UHP},
Campus Scientifique, BP 239\\
54506 Vandoeuvre-l\`es-Nancy Cedex, France}

\maketitle

\begin{abstract}
The notion of computability closure has been introduced for proving
the termination of higher-order rewriting with first-order matching by
Jean-Pierre Jouannaud and Mitsuhiro Okada in a 1997 draft which later
served as a basis for the author's PhD. In this paper, we show how
this notion can also be used for dealing with $\b$-normalized
rewriting with matching modulo $\b\eta$ (on patterns {\em\`a la}
Miller), rewriting with matching modulo some equational theory, and
higher-order data types (types with constructors having functional
recursive arguments). Finally, we show how the computability closure
can easily be turned into a reduction ordering which, in the
higher-order case, contains Jean-Pierre Jouannaud and Albert Rubio's
higher-order recursive path ordering and, in the first-order case, is
equal to the usual first-order recursive path ordering.
\end{abstract}


\section{Introduction}

After Jan Willem Klop's PhD thesis on Combinatory Reduction Systems
(CRS) \cite{klop80phd,klop93tcs}, the interest in higher-order
rewriting, or the combination of $\l$-calculus and rewriting, was
relaunched by Dale Miller and Gopalan Nadathur's work on $\l$-Prolog
\cite{miller88iclp} and Val Breazu-Tannen's paper on the modularity of
confluence for the combination of simply-typed $\l$-calculus and
first-order rewriting \cite{breazu88lics,breazu94ic}. A year later,
Dale Miller proved the decidability of unification modulo $\b\eta$ for
``higher-order patterns'' \cite{miller89elp,miller91jlc}, and the
modularity of termination for simply-typed $\l$-calculus and
first-order rewriting was independently proved by Jean Gallier and Val
Breazu-Tannen \cite{breazu89icalp,breazu91tcs} and Mitsuhiro Okada
\cite{okada89issac}, both using Jean-Yves Girard's technique of
reducibility predicates \cite{girard71sls,girard72phd,girard88book}. A
little bit later, Daniel Dougherty showed, by purely syntactic means
(without using reducibility predicates), that these results could be
extended to any ``stable'' set of untyped $\l$-terms
\cite{dougherty91rta,dougherty92ic}, the set of simply-typed
$\l$-terms being stable. We must also mention Zhurab Khasidashvili's
new approach to higher-order rewriting with his Expression Reduction
Systems (ERS) \cite{khasidashvili90}.

Then, in 1991, two important papers were published on this subject,
both introducing a new approach to higher-order rewriting: Tobias
Nipkow's Higher-order Rewrite Systems (HRS)
\cite{nipkow91lics,mayr98tcs}, and Jean-Pierre Jouannaud and Mitsuhiro
Okada's Executable Higher-Order Algebraic Specification Languages
\cite{jouannaud91lics,jouannaud97tcs}. Tobias Nipkow's approach is
based on Dale Miller's result: the simply-typed $\l$-calculus, which
is confluent and terminating, is used as a framework for encoding
higher-order rewriting. He extends to this framework the Critical Pair
Lemma. Jean-Pierre Jouannaud and Mitsuhiro Okada's approach can be
seen as a typed version of CRS's (restricted to first-order
matching). They proved that termination is modular for the combination
of simply-typed $\l$-calculus, a non-duplicating\footnote{$l\a r$ is
non-duplicating if no variable has more occurrences in $r$ than it has
in $l$.} terminating first-order rewrite system, and an higher-order
rewrite system which definition follows a ``general schema'' extending
primitive recursion. Later, Vincent van Oostrom and Femke van
Raamsdonk compared CRS's and HRS's \cite{oostrom95hoa} and developed
an axiomatized framework subsuming them
\cite{oostrom94phd,raamsdonk96phd}.

The combination of $\b$-reduction and rewriting is naturally used in
dependent type systems and proof assistants implementing the
proposition-as-type and proof-as-object paradigm
\cite{blanqui05mscs}. In these systems, two propositions equivalent
modulo $\b$-reduction and rewriting are considered as equivalent (\eg
$P(2+2)$ and $P(4)$). This is essential for enabling users to
formalize large proofs with many computations, as recently shown by
Georges Gonthier and Benjamin Werner's proof of the Four Color Theorem
in the Coq proof assistant. However, checking the correctness of user
proofs requires to check the equivalence of two terms. Hence, the
necessity to have termination criteria for the combination of
$\b$-reduction and a set $R$ of higher-order rewrite rules.

For proving the correctness of the general schema, Jean-Pierre
Jouannaud and Mitsuhiro Okada used Jean-Yves Girard's technique of
reducibility predicates. Roughly speaking, since proving the (strong)
$\b$-normalization by induction on the structure of terms does not
work directly, one needs to prove a stronger predicate. In 1967,
William Tait introduced a ``convertibility predicate'' for proving the
weak normalization of some extension of Kurt G\"odel's system T
\cite{tait67jsl}. Later, in 1971, Jean-Yves Girard introduced
``reducibility predicates'' (called {\em computability predicates} in
the following) for proving the weak and strong normalization of the
polymorphic $\l$-calculus \cite{girard71sls,girard72phd}. This
technique can be applied to (higher-order) rewriting by proving that
every function symbol is computable, that is, that every function call
is computable whenever its arguments so are.

This naturally leads to the following question: which operations
preserve computability? Indeed, from a set of such operations, one can
define the {\em computability closure} of a term $t$, written
$\CC(t)$, as the set of terms that are computable whenever $t$ so
is. Then, to get normalization, it suffices to check that, for every
rule $f\vl\a r\in R$, $r$ belongs to the computability closure of
$\vl$. Examples of computability-preserving operations are:
application, function calls on arguments smaller than $\vl$ in some
well-founded ordering $>$, etc. Jean-Pierre Jouannaud and Mitsuhiro
Okada introduced this notion in a 1997 draft which served as a basis
for \cite{blanqui99rta,blanqui02tcs}. In this paper, we show how this
notion can be extended for dealing with $\b$-normalized rewriting with
matching modulo $\b\eta$ on patterns {\em\`a la} Miller and matching
modulo some equational theory.

Another way to prove the termination of $R$ is to find a decidable
well-founded rewrite relation containing $R$. A well known such
relation in the first-order case is the recursive path ordering
\cite{plaisted78tr,dershowitz82tcs} which well-foundedness was
initially based on Kruskal theorem \cite{kruskal60ams}. The first
attempts made for generalizing this ordering to the higher-order case
were not able to orient system T
\cite{loria92ctrs,lysne95rta,jouannaud96rta}. Finally, in 1999,
Jean-Pierre Jouannaud and Albert Rubio succeeded in finding such an
ordering \cite{jouannaud99lics} by using computability-based
techniques again, hence providing the first well-foundedness proof of
RPO not based on Kruskal theorem. This ordering was later extended to
the calculus of constructions by Daria Walukiewicz
\cite{walukiewicz03phd,walukiewicz03jfp}.

Although the computability closure on one hand, and the recursive path
ordering on the other hand, share the same computability-based
techniques, there has been no precise comparison between these two
termination criteria. In \cite{walukiewicz03jfp}, one can find
examples of rules that are accepted by one criterion but not the
other. And Jean-Pierre Jouannaud and Albert Rubio themselves use the
notion of computability closure for strengthening HORPO.

In this paper, we explore the relations between both criteria. We
start from the trivial remark that the computability closure itself
provides us with an ordering: let $t~\CR(>)~u$ if $t=f\vt$ and
$u\in\CC_>(\vt)$, where $\CC_>$ is the computability closure built by
using a well-founded relation $>$ for comparing the arguments between
function calls. Proving the well-foundedness of this ordering simply
consists in proving that the computability closure is correct, which
can be done by induction on $>$. Then, we remark that the function
mapping $>$ to $\CR(>)$ is monotone wrt inclusion. Thus, it admits a
least fixpoint which is a well-founded ordering. We prove that this
fixpoint contains HORPO and is equal to RPO in the first-order case.


\section{Terms and types}
\label{sec-terms}

We consider simply-typed $\l$-terms with curried constants. See
\cite{barendregt92book} for details about typed $\l$-calculus.
For rewriting, we follow the notations of Nachum Dershowitz and
Jean-Pierre Jouannaud's survey \cite{dershowitz90book}.


Let $\cB$ be a set of {\em base types}. The set $\bT$ of {\em simple
types} is inductively defined as usual: $T\in\bT=B\in\cB~|~T\A T$.


Let $\cX$ be a set of {\em variables} and $\cF$ be a set of {\em
function symbols} disjoint from $\cX$. We assume that every
$a\in\cX\cup\cF$ is equipped with a type $\tau_a\in\bT$. The sets
$\cT^T$ of {\em terms of type $T$} are inductively defined as follows:

\begin{lst}{--}
\item If $a\in\cX\cup\cF$, then $a\in\cT^{\tau_a}$.
\item If $x\in\cX$ and $t\in\cT^U$, then $\lx t\in\cT^{\tau_x\A U}$.
\item If $v\in\cT^{T\A U}$ and $t\in\cT^T$, then $vt\in\cT^U$.
\end{lst}

As usual, we assume that, for all type $T$, the set of variables of
type $T$ is infinite and consider terms up to $\alpha$-conversion
(type-preserving renaming of bound variables). Let $\FV(t)$ be the set
of variables {\em free} in $t$. Let $\vt$ denote a sequence of terms
$t_1,\ldots,t_n$ of length $n=|\vt|\ge 0$.

Let $\tau(t)$ denote the type of a term $t$. In the following,
writing $t:T$ or $t^T$ means that $\tau(t)=T$.

The set $\pos(t)$ of positions in a term $t$ is defined as usual as
words on $\{1,2\}$. Let $t|_p$ be the subterm of $t$ at position
$p\in\pos(t)$, and $t[u]_p$ be the term obtained by replacing in $t$
its subterm at position $p\in\pos(t)$ by $u$.

A term is {\em algebraic} if it contains no abstraction and no subterm
of the form $xt$. A term $t$ is {\em linear} if no variable free in
$t$ occurs more than once in $t$.

The $\b$-reduction is the closure by context of the relation $(\lx
t)u\ab t_x^u$ where $t_x^u$ denotes the higher-order substitution of
$x$ by $u$ in $t$.


A {\em rewrite rule} is a pair of terms $l\a r$ such that $l$ is of
the form $f\vl$, $\FV(r)\sle\FV(l)$ and $\tau(l)=\tau(r)$. Given a set
$R$ of rewrite rules, let $\a_R$ be the closure by context and
substitution of $R$. Hence, matching is modulo $\alpha$-conversion
(but $\alpha$-conversion is needed only for left-hand sides having
abstractions). A rule $l\a r$ is linear (resp. algebraic) if both $l$
and $r$ are linear (resp. algebraic).


Given a relation $\a$ on terms, let $\al$, $\a^=$ and $\a^*$ be its
inverse, its reflexive closure and its reflexive and transitive
closure respectively. Let also $\a(t)=\{t'\in\cT~|~t\a t'\}$ be the
set of reducts of $t$, and $\SN(\a)$ (resp. $\SN^T(\a)$) be the set of
terms (resp. of type $T$) that are strongly normalizable wrt $\a$. Our
aim is to prove the termination (strong normalization,
well-foundedness) of ${\a}={{\ab}\cup{\a_R}}$.

Given a relation $>$, let $>\lex$, $>\mul$ and $>\prod$ respectively
denote the lexicographic, multiset and product extensions of $>$. Note
that all these extensions are well-founded whenever $>$ is
well-founded.


\section{Computability}
\label{sec-comp}

In this section, we remind the notion of computability predicate
introduced by William Tait \cite{tait67jsl,tait72lc} and extended by
Jean-Yves Girard with the notion of {\em neutral\footnote{simple in
\cite{girard72phd}.} term} \cite{girard72phd,girard88book}. Every type
is interpreted by a set of {\em computable} terms of that type. Since
computability is defined so as to imply strong normalization, the
latter is obtained by proving that every term is computable.

In the following, we assume given a set $R$ of rewrite rules.


\begin{definition}[Reducibility candidates]
\label{def-cand}
A term is {\em neutral\,} if it is of the form $x\vv$ or of the form
$(\lx t)u\vv$. Let ${\a}={{\ab}\cup{\ar}}$. A {\em reducibility
candidate for the type $T$\,} is a set $P$ of terms such that:

\begin{enumi}{}
\item $P\sle\SN^T(\a)$.
\item $P$ is stable by $\a$: $\red{P}\sle P$.
\item If $t:T$ is neutral and $\red{t}\sle P$, then $t\in P$.
\end{enumi}

\noindent
Let $\cQ_R^T$ be the set of all reducibility candidates for the type
$T$, and $\cI_R$ be the set of functions $I$ from $\cB$ to $2^\cT$
such that, for all $\fB\in\cB$, $I(\fB)\in\cQ_R^\fB$. Given an {\em
interpretation of base types} $I\in\cI_R$, we define an interpretation
$\I{T}_R^I\in\cQ_R^T$ for every type $T$ as follows:

\begin{lst}{--}
\item $\I\fB_R^I=I(\fB)$,
\item $\I{T\A U}_R^I=
\{v\in\SN^{T\A U}~|~\all t\in\I{T}_R^I,\,vt\in\I{U}_R^I\}$.
\end{lst}
\end{definition}

One can check that $\SN^T$ is a reducibility candidate for $T$.


We now check that the interpretation of a type is a reducibility
candidate.

\begin{lemma}
\label{lem-int}
If $I\in\cI_R$ then, for all type $T$, $\I{T}_R^I\in\cQ_R^T$.
\end{lemma}

\begin{proof}
We proceed by induction on $T$. The lemma is immediate for
$T\in\cB$. Assume now that $\I{T}_R^I\in\cQ_R^T$ and
$\I{U}_R^I\in\cQ_R^U$. We prove that $\I{T\A U}_R^I\in\cQ_R^{T\A U}$.

\begin{enumi}{}
\item $\I{T\A U}_R^I\sle\SN^{T\A U}$ by definition.
\item Let $v\in\I{T\A U}_R^I$, $v'\in\red{v}$ and $t\in\I{T}_R^I$.
  We must prove that $v't\in\I{U}_R^I$. This follows from the
  facts that $\I{U}_R^I\in\cQ_R^U$, $vt\in\I{U}_R^I$ and
  $v't\in\red{vt}$.
\item Let $v^{T\A U}$ be a neutral term such that
  $\red{v}\sle\I{T\A U}_R^I$ and $t\in\I{T}_R^I$. We must
  prove that $vt\in\I{U}_R^I$. Since $v$ is neutral, $vt$ is
  neutral too. Since $\I{U}_R^I\in\cQ_R^U$, it suffices to
  prove that $\red{vt}\sle\I{U}_R^I$. Since
  $\I{T}_R^I\in\cQ_R^T$, $t\in\SN$ and we can proceed by
  induction on $t$ with $\a$ as well-founded ordering. Let
  $w\in\red{vt}$. Since $v$ is neutral, either $w=v't$ with
  $v'\in\red{v}$, or $w=vt'$ with $t'\in\red{t}$. In the former case,
  $w\in\I{U}_R^I$ since $v'\in\I{T\A U}_R^I$. In the latter
  case, we conclude by induction hypothesis on $t'$.\qed
\end{enumi}
\end{proof}


Finally, we come to the definition of computability.

\begin{definition}[Computability]
Let $I$ be the base type interpretation such that $I(\fB)=\SN^\fB$. A
term $t:T$ is {\em computable} if $t\in\I{T}_R^I$.
\end{definition}

In the following, we drop the superscript $I$ in $\I{T}_R^I$.


We do not know how to prove that computability is stable by subterm
before proving that every term is computable. However, since, on base
types, computability is equivalent to strong normalization, the
subterms of base type of a computable term are computable. This is in
particular the case for the arguments of base type of a function
symbol:

\begin{definition}[Accessibility]
For all $f:\vT\A\fB$, let $\Acc(f)=\{i~|~T_i\in\cB\}$ be the set of
{\em accessible} arguments of $f$.
\end{definition}


We now prove some properties of computable terms.

\begin{lemma}[Computability properties]
\begin{enumi}{C}
\item\label{prop-comp-lam}
If $t$, $u$ and $t_x^u$ are computable, then $(\lx t)u$ is computable.

\item\label{prop-comp}
If every symbol is computable, then every term is computable.

\item\label{prop-comp-acc}
If $f\vt$ is computable and $i\in\Acc(f)$, then $t_i$ is computable.

\item\label{prop-comp-call-hd}
A term $f\vt:\fB$ is computable whenever $\vt$ are computable and
every head-reduct of $f\vt$ is computable.

\item\label{prop-comp-symb-hd}
A symbol $f:\vT\A\fB$ is computable if every head-reduct of $f\vt$ is
computable whenever $\vt:\vT$ are computable.

\item\label{prop-comp-symb-cc}
A symbol $f$ is computable if, for every rule $f\vl\a r\in R$ and
substitution $\s$, $r\s$ is computable whenever $\vl\s$ are
computable.
\end{enumi}
\end{lemma}

\begin{proof}
\begin{enumi}{C}
\item
Since $(\lx t)u$ is neutral, it suffices to prove that every reduct is
computable. We proceed by induction on $(t,u)$ with $\a\prod$ as
well-founded ordering ($t$ and $u$ are computable). Assume that $(\lx
t)u\a v$. If $v=t_x^u$, then $t'$ is computable by
assumption. Otherwise, $v=(\lx t')u$ with $t\a t'$, or $v=(\lx t)u'$
with $u\a u'$. In both cases, we can conclude by induction hypothesis.

\item
First note that the identity substitution is computable since
variables are computable (they are neutral and irreducible). We then
prove that, for every term $t$ and computable substitution $\t$, $t\t$
is computable, by induction on $t$.

\begin{lst}{--}
\item
Assume that $t=f\in\cF$. Then, $t\t=f$ is computable by assumption.
\item
Assume that $t=x\in\cX$. Then, $t\t=x\t$ is computable by assumption.
\item
Assume that $t=\lx u$. Then, $t\t=\lx u\t$. Let $v:V$ computable. We
must prove that $t\t v$ is computable. By induction hypothesis,
$u\t_x^v$ is computable. Since $u\t$ and $v$ are computable too, by
(C\ref{prop-comp-lam}), $t\t$ is computable.
\item
Assume that $t=u^{V\A T}v$. Then, $t\t=u\t v\t$. By induction
hypothesis, $u\t$ and $v\t$ are computable. Thus, $t\t$ is
computable.
\end{lst}

\item
By definition of the interpretation of base types.

\item
By definition of the interpretation of base types, it suffices to
prove that every reduct of $f\vt$ is computable. We prove it by
induction on $\vt$ with $\a\prod$ as well-founded ordering ($\vt$ are
computable). Head-reducts are computable by assumption. For
non-head-reducts, this follows by induction hypothesis.

\item
By definition of the interpretation of arrow types and
(C\ref{prop-comp-call-hd}).

\item
After (C\ref{prop-comp-symb-hd}), it suffices to prove that every
head-reduct of $f\vt$ is computable whenever $\vt$ are computable. Let
$t'$ be a head-reduct of $f\vt$. Then, there is $l\a r\in R$ and $\s$
such that $\vt=\vl\s$ and $t'=r\s$. Thus, $t'$ is computable.\qed
\end{enumi}
\end{proof}


\section{Computability closure}
\label{sec-clos}

After the properties (C\ref{prop-comp}) and
(C\ref{prop-comp-symb-cc}), we are left to prove that, for every rule
$f\vl\a r\in R$, $r\s$ is computable whenever $\vl\s$ are
computable. This naturally leads us to find a set $\CC^f(\vl)$ of
terms $t$ such that $t\s$ is computable whenever $\vl\s$ are
computable: the computability closure of $\vl$ wrt $f$.

We can include $\vl$ and close this set with computability-preserving
operations like applying a term to another or taking the accessible
argument of a function call.

We can also include variables distinct from $\FV(\vl)$ and allow
abstraction on them by strengthening the property to prove as follows:
for all $t\in\CC^f(\vl)$, $t\s$ is computable whenever $\vl\s$ are
computable and $\s$ is computable on $\FV(t)\moins\FV(l)$.

Now, to allow function calls, the idea is to introduce a precedence on
function symbols and a well-founded ordering $>$ on function
arguments.

 
So, we assume given a quasi-ordering $\ge_\cF$ on $\cF$ which strict
part ${>_\cF}={\ge_\cF\moins\le_\cF}$ is well-founded. Let
${\simeq_\cF}={\ge_\cF\cap\le_\cF}$ be its associated equivalence
relation.

We also assume that every symbol $f$ is equipped with a {\em status}
$\mr{stat}_f\in\{\mr{lex},\mr{mul}\}$, such that
$\mr{stat}_f=\mr{stat}_g$ whenever $f\simeq_\cF g$, defining how the
arguments of $f$ must be compared: lexicographically (from left to
right, or from right to left) or by multiset.

\begin{definition}[Status relation]
The {\em status relation} associated to a relation $>$ is the relation
$(f,\vt)>\stat{}(g,\vu)$ such that $f>_\cF g$ or $f\simeq_\cF g$ and
$\vt~>\stat{f}~\vu$.
\end{definition}

Note that the status relation $>\stat{}$ is well-founded whenever $>$
so is.


We now formalize the notion of computability closure.

\begin{definition}
A function $\CC$ mapping every $f^{\vT\A\fB}$ and $\vl^\vT$ to a set
of terms $\CC^f(\vl)$ is a {\em computability closure} if, for all
$f^{\vT\A\fB}$, $\vl^\vT$, $r\in\CC^f(\vl)$ and $\t$, $r\t$ is
computable whenever $\vl\t$ are computable and $\t$ is computable on
$\cX\moins\FV(\vl)$.
\end{definition}

We now check that the computability of symbols, hence the termination
of ${\ab}\cup{\ar}$ by (C\ref{prop-comp}), can be obtained by using a
computability closure.

\begin{lemma}
\label{lem-cc-fun}
If $\CC$ is a computability closure and, for all rule $f\vl\a r\in
R$, $r\in\CC^f(\vl)$, then every symbol is computable.
\end{lemma}

\begin{proof}
It follows from (C\ref{prop-comp-symb-cc}) and the fact that
$\FV(r)\sle\FV(\vl)$.\qed
\end{proof}


\begin{figure}[ht]
\centering\caption{Higher-order computability closure\label{fig-cc}}\vsp
\fbox{\begin{minipage}{11cm}
\centering\vsp
(arg)\quad $l_i\in\CC_>^f(\vl)$\\[2mm]

(decomp-symb)\quad $\cfrac{g\vu\in\CC_>^f(\vl)\quad i\in\Acc(g)}
{u_i\in\CC_>^f(\vl)}$\\[2mm]

(prec)\quad $\cfrac{f>_\cF g}
{g\in\CC_>^f(\vl)}$\\[2mm]

(call)\quad $\cfrac{f\simeq_\cF g^{\vU\A U}\quad
\vu^\vU\in\CC_>^f(\vl)\quad \vl~>^{f\vl}\stat{f}~\vu
}{g\vu\in\CC_>^f(\vl)}$\\[2mm]


(app)\quad $\cfrac{u^{V\A T}\in\CC_>^f(\vl)\quad
v^V\in\CC_>^f(\vl)}{uv\in\CC_>^f(\vl)}$\\[2mm]

(var)\quad $\cfrac{x\notin\FV(\vl)}{x\in\CC_>^f(\vl)}$\\[2mm]

(lam)\quad $\cfrac{u\in\CC_>^f(\vl)\quad x\notin\FV(\vl)}
{\lx u\in\CC_>^f(\vl)}$\vsp\end{minipage}}
\end{figure}


We now present a computability closure similar to the one introduced
in \cite{blanqui99rta,blanqui02tcs} except that the relation $>$ used
for comparing arguments in recursive calls is replaced by an abstract
family of relations $(>^l)_{l\in\cT}$. We then prove the correctness
of this abstract computability closure under some condition.

\begin{definition}[Closure-compatibility]
\label{def-clos-comp}
A relation $\succ$ is {\em closure-compatible} with a family of
relations $(>^l)_{l\in\cT}$ if, for all $l$ and $\t$, $t\t\succ u\t$
whenever $t>^l u$, $t\t$ and $u\t$ are computable, and $\t$ is
computable on $\cX\moins\FV(l)$.
\end{definition}

Note that any relation stable by substitution $>$ is
closure-compatible with itself (the constant family equal to $>$).
This is in particular the case of the restriction of the subterm
ordering $>$ defined by $t>u$ if $u$ is a subterm of $t$ and
$\FV(u)\sle\FV(t)$.

\begin{lemma}
\label{lem-cc}
Let ${>}={(>^l)_{l\in\cT}}$ be a family of relations. The function
$\CC_>$ defined in Figure \ref{fig-cc} is a computability closure
whenever there exists a well-founded relation on computable terms
$\succ$ that is closure-compatible with $>$.
\end{lemma}

\begin{proof}
We proceed by induction, first on $(f,\vl\t)$ with $\succ\stat{}$ as
well-founded ordering (H1), and second, by induction on $\CC_>^f(\vl)$
(H2).

\begin{lst}{--}
\item[(arg)]
$l_i\t$ is computable by assumption.

\item[(decomp-symb)]
By (H2), $g\vu\t$ is computable. Thus, after (C\ref{prop-comp-acc}),
$u_i\t$ is computable.

\item[(prec)]
By (H1), $g$ is computable.

\item[(call)]
By (H2), $\vu\t$ are computable. Since $\vl>^{f\vl}\stat{f}\vu$,
$\succ$ is closure-compatible with $>$, $\vl\t$ and $\vu\t$ are
computable, and $\t$ is computable on $\cX\moins\FV(\vl)$, we have
$\vl\t\succ\stat{f}\vu\t$. Therefore, by (H1), $g\vu\t$ is computable.


\item[(app)]
By (H2), $u\t$ and $v\t$ are computable.  Thus, $u\t v\t$ is
computable.

\item[(var)]
Since $x\in\cX\moins\FV(\vl)$, $x\t$ is computable by assumption.

\item[(lam)]
Wlog we can assume that $x\notin\codom(\t)$. Thus, $(\lx u)\t=\lx
u\t$. Let $v:\tau_x$ computable. After (C\ref{prop-comp-lam}), $(\lx
u\t)v$ is computable if $u\t$, $v$ and $u\t_x^v$ are computable. We
have $v$ computable by assumption and $u\t$ and $u\t_x^v$ computable
by (H2).\qed
\end{lst}
\end{proof}


\section{$\b$-normalized rewriting with matching modulo $\b\eta$}
\label{sec-hopm}

In this section, we show how the notion of computability closure can
be extended to deal with HRS's \cite{nipkow91lics}. This extends our
previous results on CRS's and HRS's \cite{blanqui00rta}. This
computability closure approach seems simpler than the technique of
``neutralization'' introduced by Jean-Pierre Jouannaud and Albert
Rubio in \cite{jouannaud06rta-termin}. However, the comparison between
both approaches remains to be done.

In HRS's, rewrite rules are of base type, rule left-hand sides are
patterns {\em\`a la} Miller \cite{miller91jlc}, and rewriting is
defined on terms in $\b$-normal $\eta$-long form as follows: $t\A_R u$
if there are $p\in\pos(t)$, $l\a r\in R$ and $\s$ in $\b$-normal
$\eta$-long form such that ${t|_p}={l\s\!\ad_\b\au_\eta}$ and
$u=t[r\s\!\ad_\b\au_\eta]_p$.

We are going to consider a slightly more general notion of rewriting:
{\em $\b$-normalized rewriting with matching modulo $\b\eta$}, defined
as follows: $t\arbe u$ if there are $p\in\pos(t)$, $l\a r\in R$ and
$\s$ in $\b$-normal form such that $t|_p$ is in $\b$-normal form,
$t|_p\eqbe l\s$ and $u=t[r\s]_p$. Furthermore, we do not assume that
rules are of base type. However, in this case, one can check that, on
terms in $\b$-normal $\eta$-long form, ${\A_R}\sle{\arbe\ab^*}$.

Matching modulo $\b\eta$ is necessary when a rule left-hand side
contains abstractions. Consider for instance the left-hand side
$l=D\lx(\sin(Fx))$. With matching modulo $\alpha$-conversion only, the
term $t=D\lx(\sin u)$ matches $p$ only if $u$ is of the form $vx$. In
particular, $D\lx(\sin x)$ does not match $p$. Yet, if one substitutes
$F$ by $\lx u$ in $l$, then one gets $D(\lx(\sin((\lx u)x)))$ which
$\b$-reduces to $t$.

Take now $l=D\lx(Fx)$. With matching modulo $\alpha$-conversion only,
the term $t=Du$ matches $l$ only if $u$ is of the form $\lx v$. In
particular, $(D\sin)$ does not match $l$. Yet, if one substitutes $F$
by $u$ in $l$, then one gets $D\lx(ux)$ which $\eta$-reduces to $t$
since $x\notin\FV(u)$ (by definition of higher-order substitution).

Higher-order patterns are terms in $\b$-normal $\eta$-long form which
free variables are applied to terms $\eta$-equivalent to distinct
bound variables. Hence, if $l$ is a pattern, $t$ and $\s$ are in
$\b$-normal form and $l\s\eqbe t$, then $l\s\a_{\b_0}^*=_\eta t$,
where $\abz$ is the restriction of $\ab$ to redexes of the form $(\lx
t)x$, that is, $(\lx t)x\abz t$ \cite{miller91jlc}.

Now, for proving the termination of ${\ab}\cup{\arbe}$ (hence the
termination of the HRS rewrite relation $\A_R$), it suffices to adapt
the notion of computability by replacing $\ar$ by $\arbe$. One can
check that all the proofs of the computability properties are still
valid except the one for (C\ref{prop-comp-symb-cc}) for which we give
a new proof:


\begin{lemma}[C\ref{prop-comp-symb-cc}]
A symbol $f$ is computable if, for every rule $f\vl\a r\in R$ and
substitution $\s$, $r\s$ is computable whenever $\vl\s$ are
computable.
\end{lemma}

\begin{proof}
After (C\ref{prop-comp-symb-hd}), for proving that $f:\vT\A\fB$ is
computable, it suffices to prove that every head-reduct of $f\vt$ is
computable whenever $\vt:\vT$ are computable. Let $t'$ be a
head-reduct of $f\vt$. Then, $f\vt$ is in $\b$-normal form and there
are $f\vl\a r\in R$ and $\s$ such that $f\vl\s\al_{\b_0}^*=_\eta f\vt$
and $t'=r\s$. To conclude, it suffices to check that $\vl\s$ are
computable.

To this end, we prove that computability is preserved by
$\eta$-reduction, $\eta$-expansion and $\b_0$-expansion. Let $t$ be a
computable term and let $u$ be a term obtained from $t$ by
$\eta$-reduction, $\eta$-expansion or $\b_0$-expansion. We prove that
$u$ is computable when $u$ is of base type. If $u$ is not of base type
then, by applying it to computable terms of appropriate types, we get a
term of base type. On base types, computability is equivalent to
strong normalization. Thus, it suffices to prove that every reduct of
$u$ is strongly normalizable. In each case, we proceed by induction on
$t$ with $\a$ as well-founded ordering ($t$ is computable).

\begin{lst}{--}
\item $\b_0$-expansion: $t\al_{\b_0}u$.
If $u\ab u'$ then either $u'=t$ is computable or, by confluence of
$\b$ and since $\b_0$ makes no duplication, there is $t'$ such that
$t\ab t'\al_{\b_0}^* u'$. Now, if $u\ar u'$ then, since $R$-redexes
are in $\b$-normal form, the $\b_0$-redex is either above the
$R$-redex or at a disjoint position. Thus, there is $u'$ such that
$t\ar t'\al_{\b 0} u'$. In both cases, we can conclude by induction
hypothesis.

\item $\eta$-reduction: $t\a_\eta u$.
If $u\ab u'$ then, by postponement of $\eta$ wrt $\b$
(${\a_\eta\ab}\sle{\ab^+\a_\eta^*}$), there is $t'$ such that $t\ab^+
t'\a_\eta^* u'$. Now, if $u\ar u'$ then, since $R$-redexes are in
$\b$-normal form, either the $\eta$-redex is a $\b$-redex and $t\ab
u\ar t'=u'$, or there is $t'$ such that $t\ar t'\a_\eta^* u'$. In both
cases, we can conclude by induction hypothesis.

\item $\eta$-expansion: $t\al_\eta u$.
If $u\ab u'$ then either $u'=t$ is computable or, by confluence of
$\b\eta$, there is $t'$ such that $t\ab t'\al_\eta^* u'$. Now, if
$u\ar u'$ then, since $R$-redexes are in $\b$-normal form, there is
$t'$ such that $t\ar t'\al_\eta^* u'$. In both cases, we can conclude
by induction hypothesis.\qed
\end{lst}
\end{proof}

By property (C\ref{prop-comp}) and Lemma \ref{lem-cc}, it follows that
${\a}={{\ab}\cup{\arbe}}$ is well-founded if, for all rule $f\vl\a
r\in R$, $r\in\CC_>^f(\vl)$.


\begin{figure}[ht]
\centering\caption{Decomposition rules for higher-order patterns\label{fig-hopm}}\vsp
\fbox{\begin{minipage}{11cm}
\centering\vsp
(decomp-lam)\quad $\cfrac{\ly u\in\CC_>^f(\vl)\quad y\notin\FV(\vl)}
{u\in\CC_>^f(\vl)}$\\[2mm]

(decomp-app-left)\quad $\cfrac{uy\in\CC_>^f(\vl)\quad
y\notin\FV(\vl)\cup\FV(u)}{u\in\CC_>^f(\vl)}$
\vsp\end{minipage}}
\end{figure}


Now, for dealing with patterns {\em \`a la} Miller, we also need to
add new decomposition rules in the computability closure.

\begin{lemma}
The function $\CC_>$ defined by the rules of Figure \ref{fig-cc} and
\ref{fig-hopm} is a computability closure whenever there exists a
well-founded relation on computable terms that is closure-compatible
with $>$.
\end{lemma}

\begin{proof}
We extend the proof of Lemma \ref{lem-cc} with the new decomposition
rules.

\begin{lst}{--}
\item[(decomp-lam)]
Let $\t'$ be the restriction of $\t$ to $\dom(\t)\moins\{y\}$. Wlog,
we can assume that $y\notin\codom(\t)$. Hence, $(\ly u)\t'=\ly
u\t'$. Now, since $\dom(\t)\sle\FV(u)\moins\FV(\vl)$,
$\dom(\t')\sle\FV(\ly u)\moins\FV(\vl)$. Thus, by (H2), $\ly u\t'$ is
computable. Since $y\t$ is computable, $(\ly u\t')y\t$ is
computable. Thus, by $\b$-reduction, $u\t'{}_y^{y\t}$ is computable
too. Finally, since $y\notin\dom(\t')\cup\codom(\t')$,
$u\t'{}_y^{y\t}=u\t$.

\item[(decomp-app-left)]
Let $v:\tau_y$ computable. Since $\dom(\t)\sle\FV(u)\moins\FV(\vl)$ and
$y\notin\FV(\vl)$,
$\dom(\t_y^v)=\dom(\t)\cup\{y\}\sle\FV(uy)\moins\FV(\vl)$. Thus, by
(H2), $(uy)\t_y^v=u\t_y^vv$ is computable. Since $y\notin\FV(u)$,
$u\t_y^v=u\t$. Thus, $u\t$ is computable.\qed
\end{lst}
\end{proof}


\section{Matching modulo some equational theory}
\label{sec-ac}

In this section, we show how the notion of computability closure can
be used for proving the termination of the combination of
$\b$-reduction and rewriting with matching modulo some equational
theory $E$ \cite{peterson81jacm,jouannaud86siam}.

To this end, we assume that $E$ is a symmetric set of {\em rules},
that is, $l\a r\in E$ iff $r\a l$ in $E$. By definition of rewrite
rules (see Section \ref{sec-terms}), this implies that, for all $l\a
r\in E$, $r$ is of the form $g\vr$ and $\FV(l)=\FV(r)$. This includes
associativity and commutativity but excludes collapsing rules like
$x+0\a x$ and erasing rules like $x\times 0\a 0$.

Then, rewriting with matching modulo can be defined as follow: $t\are
u$ if there are $p\in\pos(t)$, $l\a r\in R$ and $\s$ such that
$t|_p\ae^* l\s$ and $u=t[r\s]_p$.

Rewriting with matching modulo $E$ is different from rewriting modulo
$E$ which is $\ae^*\ar$. The point is that, with matching modulo $E$,
no $E$-step takes place above $t|_p$ when one rewrites a term $t$ at
some position $p\in\pos(t)$.

Hence, we correct an error in \cite{blanqui03rta} (Theorem 6) where it
is claimed that ${\ab}\cup{\ae^*\ar}$ is terminating. What is in fact
proved in \cite{blanqui03rta} is the termination of
${\ab}\cup{\a_{E_1}^*\a_{R_1}}\cup{\a_{R_\w,E_\w}}$ where $E_1$ and
$R_1$ (resp. $E_\w$ and $R_\w$) are the first-order
(resp. higher-order) parts of $E$ and $R$ respectively.

For proving the termination of ${\ab}\cup{\are}$, it suffices to adapt
computability by replacing $\ar$ by $\are$. One can check that all the
proofs of computability properties are still valid except the one for
(C\ref{prop-comp-symb-cc}) for which we give a new proof:

\begin{lemma}[C\ref{prop-comp-symb-cc}]
Let $E$ be a symmetric set of {\em rules}. Assume that $\succ$ is a
well-founded relation on computable terms closure-compatible with $>$
and that, for all rule $f\vl\a g\vr\in E$, $\vr\in\CC_>^f(\vl)$.
Then, $f$ is computable if, for every rule $f\vl\a r\in R$ and
substitution $\s$, $r\s$ is computable whenever $\vl\s$ are
computable.
\end{lemma}

\begin{proof}
By Lemma \ref{lem-cc}, $\CC_>$ is a computability closure. After
(C\ref{prop-comp-symb-hd}), for proving that $f:\vT\A\fB$ is
computable, it suffices to prove that every head-reduct of $f\vt$ is
computable whenever $\vt:\vT$ are computable. Let $t'$ be a
head-reduct of $f\vt$. Then, there is $g\vl\a r\in R$ and $\s$ such
that $f\vt\ae^*g\vl\s$ and $t'=r\s$. By definition of computability
closure, $\vl\s$ are computable since $\vt$ are computable (induction
on the number of $E$-steps). Therefore, $r\s$ is computable.\qed
\end{proof}

By property (C\ref{prop-comp}) and Lemma \ref{lem-cc}, it follows that
${\a}={{\ab}\cup{\are}}$ is well-founded if moreover, for all rule
$f\vl\a r\in R$, $r\in\CC_>^f(\vl)$.


\section{Higher-order data types}
\label{sec-hodt}

Until now, we used the subterm ordering in (call). But this ordering
is not strong enough to handle recursive definitions on higher-order
data types, \ie data types with constructors having functional
recursive arguments. Consider for instance a type $\fP$ representing
processes with a sequence operator $;:\fP\A\fP\A\fP$ and a
data-dependent choice operator $\S:(\fD\A\fP)\A\fP$. Then, in the
following simplification rule \cite{vandepol93hoa}:

\begin{center}
$(\S P);x\a \S\ly(Py;x)$
\end{center}

\noindent
the term $Py$ is not a subterm of $\S P$.

In this section, we describe an extension of the computability closure
to handle such definitions. It is based on the interpretation of
``positive'' higher-order data types introduced by Nax Paul Mendler in
1987 \cite{mendler87lics,mendler91pal}.

As usual, the set $\pos(T)$ of {\em positions in a type $T$} is
defined as words on $\{1,2\}$. The sets $\pos^+(T)$ and $\pos^-(T)$ of
{\em positive and negative positions} respectively are inductively
defined as follows:

\begin{lst}{--}
\item $\pos^\d(\fB)=\{\vep\}$.
\item $\pos^\d(T\A U)= 1\cdot\pos^{-\d}(T)\cup 2\cdot\pos^\d(U)$.
\end{lst}

Let $\pos(\fB,T)$ be the positions of the occurrences of $\fB$ in
$T$. A base type $\fB$ {\em occurs only positively} (resp. {\em
negatively}) in a type $T$ if $\pos(\fB,T)\sle\pos^+(T)$
(resp. $\pos(\fB,T)\sle\pos^-(T)$).

Nax Paul Mendler showed that the combination of $\b$-reduction and
reduction rules for a ``case'' or ``match'' construction does not
terminate if a data type $\fB$ has a constructor having an argument in
the type of which $\fB$ occurs negatively (we say that $\fB$ is not
positive). Take for instance $c:(\fB\A\fN)\A\fB$, $f:\fB\A(\fB\A\fN)$
together with the rule $f(cx)\ar x$. Then, by taking $\w=\lx
fxx:\fB\A\fN$, we have $\w(c\w)\ab f(c\w)(c\w)\ar \w(c\w)\ab \ldots$

He also showed that the set of all reducibility candidates is a
complete lattice for inclusion and that, if $\fB$ is positive, then
one can build an interpretation of $\fB$ as the fixpoint of a monotone
functional on reducibility candidates, in which the reduction rules
for the case construction are safe. In this case, we can say that
every argument of a constructor is accessible. We extend this notion
of accessibility to every (defined or undefined) function symbol as
follows.


\begin{definition}[Accessible arguments]
For every $f^{\vT\A\fB}\in\cF$, let $\Acc(f)=
\{i\le|\vT|~|~ \pos(\fB,T_i)\sle\pos^+(T_i)\}$.
\end{definition}

In our example, we have $\pos(\fP,\fD\A\fP)=\{2\}=\pos^+(\fD\A\fP)$
and $\pos(\fP,\fP)=\{\vep\}=\pos^+(\fP)$. Thus, $\Acc(\S)=\{1\}$ and
$\Acc(;)=\{1,2\}$.


We now define the functional the least fixpoint of which will provide
the interpretation of base types.

\begin{lemma}
The function $F_R^I(\fB)=\{t\in\SN^B~|~\all f^{\vT\A\fB}\vt,\,t\a^*
f\vt\A\all i\in\Acc(f),\,t_i\in\I{T_i}_R^I\}$ is a monotone function
on $\cI_R$.
\end{lemma}

\begin{proof}
We first prove that $P=F_R^I(\fB)\in\cQ_R^\fB$.

\begin{enumi}{}
\item $P\sle\SN^\fB$ by definition.
\item Let $t\in P$, $t'\in\red{t}$, $f:\vT\A\fB$ and $\vt$
  such that $t'\a^* f\vt$. We must prove that
  $\vt\in\I{\vT}_R$. It follows from the facts that $t\in P$ and
  $t\a^* f\vt$.
\item Let $t^\fB$ neutral such that $\red{t}\sle P$.
  Let $f^{\vT\A\fB}$, $\vt$ such that $t\a^* f\vt$ and
  $i\in\Acc(f)$. We must prove that $t_i\in\I{T_i}_R$. Since $t$ is
  neutral, $t\neq f\vt$. Thus, there is $t'\in\red{t}$ such that
  $t'\a^* f\vt$. Since $t'\in P$, $t_i\in\I{T_i}_R$.
\end{enumi}

For the monotony, let ${\le^+}={\le}$ and ${\le^-}={\ge}$. Let $I\le
J$ iff, for all $\fB$, $I(\fB)\sle J(\fB)$. We first prove that
$\I{T}_R^I\sle^\d \I{T}_R^J$ whenever $I\le J$ and
$\pos(\fB,T)\sle\pos^\d(T)$, by induction on $T$.

\begin{lst}{--}
\item Assume that $T=C\in\cB$. Then, $\d=+$, $\I{T}_R^I=I(C)$
  and $\I{T}_R^I=J(C)$. Since $I(C)\sle J(C)$, $\I{T}_R^I\sle
  \I{T}_R^I$.
\item Assume that $T=U\A V$. Then, $\pos(\fB,U)\sle\pos^{-\d}(U)$ and
  $\pos(\fB,V)\sle\pos^\d(V)$. Thus, by induction hypothesis,
  $\I{U}_R^I\sle^{-\d} \I{U}_R^J$ and $\I{V}_R^I\sle^\d
  \I{V}_R^J$. Assume that $\d=+$. Let $t\in\I{T}_R^I$ and
  $u\in\I{U}_R^J$. We must prove that $tu\in\I{V}_R^J$. Since
  $\I{U}_R^I\sge \I{U}_R^J$, $tu\in\I{V}_R^I$. Since $\I{V}_R^I\sle
  \I{V}_R^J$, $tu\in\I{V}_R^J$. It works similarly for $\d=-$.
\end{lst}

Assume now that $I\le J$. We must prove that, for all $\fB$,
$F_R^I(\fB)\sle F_R^J(\fB)$. Let $\fB\in\cB$ and $t\in F_R^I(\fB)$. We must
prove that $t\in F_R^J(\fB)$. First, we have $t\in\SN^\fB$ since $t\in
F_R^I(\fB)$. Assume now that $t\a^* f^{\vT\A\fB}\vt$ and let
$i\in\Acc(f)$. We must prove that $t_i\in\I{T_i}_R^J$. Since $t\in
F_R^I(\fB)$, $t_i\in\I{T_i}_R^I$. Since $i\in\Acc(f)$,
$\pos(\fB,T_i)\sle\pos^+(T_i)$ and $\I{T_i}_R^I\sle \I{T_i}_R^J$.\qed
\end{proof}


\begin{definition}[Computability]
Let $I_R$ be the least fixpoint of $F_R$. A term $t:T$ is {\em
computable} if $t\in\I{T}_R^{I_R}$.
\end{definition}

In the following, we drop the superscript $I_R$ in $\I{T}_R^{I_R}$.

One can check that all the proofs of computability properties are
still valid except the one for (C\ref{prop-comp-call-hd}) for which we
give a new proof:

\begin{lemma}[C\ref{prop-comp-call-hd}]
A term $f\vt:\fB$ is computable whenever $\vt$ are computable and
every head-reduct of $f\vt$ is computable.
\end{lemma}

\begin{proof}
We first need to prove that $f\vt$ is $\SN$. This follows from the
previous proof of (C\ref{prop-comp-call-hd}). Assume now that
$f\vt\a^*g\vu$ and $i\in\Acc(g)$. We prove that $u_i$ is computable by
induction on $\vt$ with $\a\prod$ as well-founded ordering ($\vt$ are
computable). If $f\vt=g\vu$, then $u_i=t_i$ is computable by
assumption. Otherwise, $f\vt\a v\a^*g\vu$. If $v$ is a head-reduct of
$f\vt$, then $v$ and $u_i$ are computable. Otherwise, we conclude by
induction hypothesis.\qed
\end{proof}


The least fixpoint of $F_R$ is reachable by transfinite iteration from
the smallest element of $\cI_R$. This provides us with an ordering
that can handle definitions on higher-order data types.

\begin{definition}[Size ordering]
For all $\fB\in\cB$ and $t\in\I\fB_R$, let the {\em size} of $t$ be
the smallest ordinal $o_R^\fB(t)=\ka$ such that $t\in
F_R^\ka(\vide)(\fB)$, where $F_R^\ka$ is the transfinite
$\ka$-iteration of $F_R$. Let $\succeq_R$ be the union of all the
relations $\succeq_R^T$ inductively defined on $\I{T}_R$ as follows:

\begin{lst}{--}
\item $t\succeq_R^\fB u$ if $o_R^\fB(t)\ge o_R^\fB(u)$.
\item $t\succeq_R^{T\A U} u$ if, for all $v\in\I{T}_R$, $tv\succeq_R^U uv$. 
\end{lst}
\end{definition}

In our example, we have $\I\fP_R=\{t\in\SN^P~|~\all
f^{\vT\A\fP}\vt,\,t\a^* f\vt\A\all
i\in\Acc(f),\,t_i\in\I{T_i}_R\}$. Since $\Acc(\S)=\{1\}$, if $\S
P\in\I\fP_R$ then, for all $d\in\I\fD_R$, $Pd\in\I\fP_R$ and
$o_R^\fP(Pd)<o_R^\fP(\S P)$.


We immediately check that the size ordering is well-founded.

\begin{lemma}
$\succeq_R$ is a well-founded quasi-ordering containing $\a$.
\end{lemma}

\begin{proof}
The relation $\succeq_R$ is the union of pairwise disjoint
relations. Hence, it suffices to prove that each one is transitive and
well-founded. We proceed by induction on $T$. For $T\in\cB$, this is
immediate. Assume now that $(t_i)_{i\in\bN}$ is an increasing sequence
for $\succ_R^{T\A U}$. Since variables are computable, let
$x\in\I{T}_R$. By definition of $\succ_R^{T\A U}$, $(t_ix)_{i\in\bN}$
is an increasing sequence for $\succ_R^U$.\qed
\end{proof}


\begin{figure}[ht]
\centering\caption{Accessibility ordering\label{fig-acc-ord}}\vsp
\fbox{\begin{minipage}{11cm}
\centering\vsp
($>$base)\quad $\cfrac{i\in\Acc(g)\quad
\vb\in\cX\moins\FV(l)
}{g^{\vA\A B}\va^\vA>^l a_i^{\vB\A B}\vb^\vB}$\\[2mm]

($>$lam)\quad $\cfrac{a>^l bx
\quad x\in\cX\moins (\FV(b)\cup\FV(l))}
{\lx a>b}$\\[2mm]

($>$red)\quad $\cfrac{a>^l b\quad b\ab c}{a>^l c}$\\[2mm]

($>$trans)\quad $\cfrac{a>^l b\quad b>^l c}
{a>^l c}$\vsp
\end{minipage}}
\end{figure}


We now define some relation strong enough for capturing definitions on
higher-order data types and with which $\succ_R$ is closure-compatible.

\begin{lemma}
$\succ_R$ is closure-compatible with the family $(>^l)_{l\in\cT}$
defined Figure \ref{fig-acc-ord}.
\end{lemma}

\begin{proof}
We prove that $a\t\succ_R b\t$ whenever $a>^l b$, $a\t$ and $b\t$ are
computable, and $\t$ is computable on $\cX\moins\FV(l)$.

\begin{lst}{--}
\item[($>$base)]
By definition of $I_R$, $o_R(g\va\t)=\ka+1$ and
$a_i\t\in\I{\vB\A\fB}_R^{I_R^\ka}$. Since $\vb\in\cX\moins\FV(l)$ and
$\t$ is computable on $\cX\moins\FV(l)$, $\vb\t$ are
computable. Therefore, $a_i\t\vb\t\in I_R^\ka(B)$ and
$a_R(g\va\t)>\ka\ge o_R(a_i\t\vb\t)$.

\item[($>$lam)]
Let $w:\tau_x$ computable. Wlog we can assume that
$x\notin\dom(\t)\cup\codom(\t)$. Hence, $(\lx a)\t=\lx a\t$. We must
prove that $(\lx a\t)w\succ_R b\t w$. By $\b$-reduction, $(\lx
a\t)w\succeq_R a\t_x^w$. By induction hypothesis, $a\t_x^w\succ_R
(bx)\t_x^w$. Since $x\notin\FV(b)\cup\dom(\t)\cup\codom(\t)$,
$(bx)\t_x^w=b\t w$.

\item[($>$red)]
By induction hypothesis and since ${\ab}\sle{\succeq_R}$.

\item[($>$trans)] By induction hypothesis and transitivity of $\succ_R$.\qed
\end{lst}
\end{proof}

By property (C\ref{prop-comp}) and Lemma \ref{lem-cc}, it follows that
${\a}={{\ab}\cup{\ar}}$ is well-founded if, for all rule $f\vl\a r\in
R$, $r\in\CC_>^f(\vl)$.

Note that we could strengthen the definition of $(>^l)_{l\in\cT}$ by
taking in ($>$base), when $l=f\vl$, $\vb\in\CC_>^f(\vl)$ instead of
$\vb\in\cX\moins\FV(\vl)$, making the definitions of $>$ and $\CC_>$
mutually dependent. See \cite{blanqui06rr} for details.


\section{The recursive computability ordering}

We now show how the computability closure can be turned into a
well-founded ordering containing the monomorphic version of
Jean-Pierre Jouannaud and Albert Rubio's higher-order recursive path
ordering \cite{jouannaud99lics}.

Indeed, consider the relation
$\CR(>)=\{(f\vl,r)~|~r\in\CC_>^f(\vl),\FV(r)\sle\FV(\vl),\tau(f\vl)=\tau(r)\}$
made of all the rules which right-hand side is in the computability
closure of its left-hand side. After (C\ref{prop-comp}) and Lemma
\ref{lem-cc-fun}, ${\ab}\cup{\a_{\CR(>)}}$ is well-founded whenever $>$
is well-founded and stable by substitution. Hence, $\CR(>)$ is itself
well-founded and stable by substitution whenever $>$ is well-founded
and stable by substitution.

We now observe that the function mapping $>$ to $\CR(>)$ is monotone
wrt inclusion. It has therefore a least fixpoint that is stable by
substitution and which closure by context is well-founded when
combined with $\ab$.

\begin{lemma}
The function mapping $>$ to the relation
$\CR(>)=\{(f\vl,r)~|~r\in\CC_>^f(\vl)$, $\FV(r)\sle\FV(\vl)$,
$\tau(f\vl)=\tau(r)\}$ is monotone wrt inclusion on the set of
well-founded relations stable by substitution.
\end{lemma}

\begin{proof}
Assume that ${>_1}\sle{>_2}$. One can prove by induction on
$(f\vl,r)\in\CR(>_1)$ that $(f\vl,r)\in\CR(>_2)$. In the (call) case,
we use the fact that the function mapping $>$ to $>\stat{}$ is
monotone wrt inclusion.

Now, assume that $>$ is well-founded and stable by substitution. After
(C\ref{prop-comp}) and Lemma \ref{lem-cc-fun},
${\ab}\cup{\a_{\CR(>)}}$ is well-founded. Thus, $\CR(>)$ is
well-founded. Now, one can check that $\CR(>)$ is stable by
substitution whenever $>$ is stable by substitution.\qed
\end{proof}

\begin{definition}
Let the {\em weak higher-order recursive computability (quasi-)
ordering} $>\whorco$ be the least fixpoint of $\CR$, and the {\em
higher-order recursive computability (quasi-) ordering} $>\horco$ be
the closure by context of $>\whorco$.
\end{definition}

In Figure \ref{fig-rco}, we give an inductive presentation of
$>\horco$ obtained by replacing $u\in\CC_>^f(\vl)$ by $f\vl>u$ in
Figure \ref{fig-cc}, and adding a rule (cont) for the closure by
context and a rule (rule) for the conditions on rules.

Strictly speaking, $>\horco$, like $>\horpo$, is not a
quasi-ordering. One needs to take its transitive closure to get a
quasi-ordering. On the other hand, one can check that $>\whorco$ is
transitive, hence is a true quasi-ordering (note that, if $t>\whorco
u$, then $t$ is of the form $f\vt$).

Moreover, since $>\whorco$ is not closed by context, it is better
suited for proving the termination of rewrite systems by using the
dependency pair method \cite{arts00tcs,sakai01rpc,blanqui06wst-hodp}.


\begin{figure}[ht]
\centering\caption{Higher-order computability ordering\label{fig-rco}}\vsp
\fbox{\begin{minipage}{11cm}
\centering\vsp
(cont)\quad $\cfrac{t>\whorco u\quad p\in\pos(C)}
{C[t]_p>\horco C[u]_p}$\\[2mm]

(rule)\quad $\cfrac{t^T>u^U\quad \FV(u)\sle\FV(t)\quad T=U}
{t>\whorco u}$\\[2mm]

(arg)\quad $f\vl>l_i$\\[2mm]

(decomp-symb)\quad $\cfrac{f\vl>g\vu\quad i\in\Acc(g)}{f\vl>u_i}$\\[2mm]

(prec)\quad $\cfrac{f>_\cF g}{f\vl>g}$\\[2mm]

(call)\quad $\cfrac{f\simeq_\cF g^{\vU\A U}\quad
f\vl>\vu^\vU\quad \vl~(>\whorco)\stat{f}~\vu
}{f\vl>g\vu}$\\[2mm]


(app)\quad $\cfrac{f\vl>u^{V\A T}\quad f\vl>v^V}{f\vl>uv}$\\[2mm]

(var)\quad $\cfrac{x\notin\FV(\vl)}{f\vl>x}$\\[2mm]

(lam)\quad $\cfrac{f\vl>u\quad x\notin\FV(\vl)}{f\vl>\lx u}$
\vsp\end{minipage}}
\end{figure}


\begin{figure}[ht]
\centering\caption{HORPO \cite{jouannaud99lics}\label{fig-horpo}}\vsp
\fbox{\begin{minipage}{11cm}
\centering\vsp
$P(f,\vt,u)= f\vt>\horpo u\ou (\ex j)~ t_j\ge\horpo u$
\vsp

(1)\quad $\cfrac{t_i\ge\horpo u}{f^{\vT\A T}\vt^\vT>\horpo u^T}$\\[2mm]

(2)\quad $\cfrac{f>_\cF g\quad P(f,\vt,\vu)}
{f^{\vT\A T}\vt^\vT>\horpo g^{\vU\A T}\vu^\vU}$\\[2mm]

(3)\quad $\cfrac{f\simeq_\cF g\quad
\mr{stat}_f=\mr{mul}\quad \vt~(>\horpo)\stat{f}~\vu}
{f^{\vT\A T}\vt^\vT>\horpo g^{\vU\A T}\vu^\vU}$\\[2mm]

(4)\quad $\cfrac{f\simeq_\cF g\quad \mr{stat}_f=\mr{lex}\quad
\vt~(>\horpo)\stat{f}~\vu\quad P(f,\vt,\vu)}
{f^{\vT\A T}\vt^\vT>\horpo g^{\vU\A T}\vu^\vU}$\\[2mm]

(5)\quad $\cfrac{P(f,\vt,\vu)}{f^{\vT\A T}\vt>\horpo \vu^T}$\\[2mm]

(6)\quad $\cfrac{\{t_1,t_2\}~(>\horpo)\mul~ \{u_1,u_2\}}
{t_1^{U\A T}t_2^U>\horpo u_1^{V\A T}u_2^V}$\\[2mm]

(7)\quad $\cfrac{t>\horpo u}{\lx t>\horpo \lx u}$\vsp
\end{minipage}}
\end{figure}


We now would like to compare this ordering with the monomorphic
version of $>\horpo$ which definition is reminded in Figure
\ref{fig-horpo}. To this end, we need to slightly strengthen the
definition of computability closure by replacing $>$ by its closure by
context $\a_>$, and by adding the following deduction rule:

\begin{center}
(red)\quad $\cfrac{u\in\CC_>^f(\vl)\quad u>v}{v\in\CC_>^f(\vl)}$
\end{center}

\noindent
One can check that all the properties are preserved. More details can
be found in \cite{blanqui06rr}. Hence, we get the following additional
deduction rules for $>\whorco$:

\vsp\fbox{
\begin{minipage}{10.75cm}
\begin{center}
(call)\quad $\cfrac{f\simeq_\cF g^{\vU\A U}\quad
f\vl>\vu^\vU\quad \vl~(>\horco)\stat{f}~\vu
}{f\vl>g\vu}$\\[2mm]

(red)\quad $\cfrac{f\vl>u\quad u>\horco v}{f\vl>v}$
\end{center}
\end{minipage}
}\vsp


We now prove that $>\horpo$ is included in the transitive closure of
$>\horco$.

\begin{lemma}
${>\horpo}\sle{>\horco^+}$.
\end{lemma}

\begin{proof}
Note that $\FV(u)\sle\FV(t)$ and $T=U$ whenever $t^T>\horpo u^U$
($>\horpo$ is a set of rules).

We first prove the property (*): $f\vt>v$ whenever $t_j>\horco^* v$ or
$f\vt>\horco^+ v$. Assume that $t_j>\horco^* v$. By (arg),
$f\vt>t_j$. Thus, by (red), $f\vt>v$. Assume now that $f\vt>\horco
u>\horco^* v$. By (red), it suffices to prove that $f\vt>u$. There are
two cases:

\begin{lst}{--}
\item $f\vt=f\va t_k\vb$, $u=f\va t_k'\vb$ and $t_k>\horco t_k'$.
We conclude by (call).

\item $f\vt=f\vl\vb$, $u=r\vb$ and $f\vl>\whorco r$.
One can check that $f\vl t>rt$ whenever $f\vl>r$.
\end{lst}

We now prove the theorem by induction on $>\horpo$.

\begin{enumi}{}
\item By induction hypothesis, $t_i>\horco^* u$. By (arg), $f\vt>t_i$.
Since $t_i>\horpo u$ and $f\vt>\horpo u$, $f\vt\a t_i$ is a
rule. Thus, $f\vt>\whorco t_i$ and, by (red), $f\vt>\whorco u$.

\item By induction hypothesis, for all $i$, $f\vt>\horco^+ u_i$ or
$t_j>\horco^* u_i$. Hence, by (*), $f\vt>\vu$. By (prec),
$f\vt>g$. Thus, by (app), $f\vt>g\vu$. Since $f\vt\a g\vu$ is a rule,
$f\vt>\whorco g\vu$.

\item By induction hypothesis, $\vt~(>\horco^+)\mul~\vu$.
Hence, by (*), $f\vt>\vu$. Thus, by (call), $f\vt>g\vu$. Since $f\vt\a
g\vu$ is a rule, $f\vt>\whorco g\vu$.

\item By induction hypothesis, $\vt~(>\horco^+)\stat{f}~\vu$ and,
for all $i$, $f\vt>\horco^+ u_i$ or $t_j>\horco^* u_i$. Hence, by (*),
$f\vt>\vu$. Thus, by (call), $f\vt>g\vu$. Since $f\vt\a g\vu$ is a
rule, $f\vt>\whorco g\vu$.

\item By induction hypothesis, for all $i$, $f\vt>\horco^+ u_i$ or
$t_j>\horco^* u_i$. Hence, by (*), $f\vt>u_i$ for all $i$. Thus, by (app),
$f\vt>\vu$. Since $(f\vt,\vu)$ is a rule, $f\vt>\whorco\vu$.

\item For typing reasons, $(t_1,u1)~(>\horpo)\prod~(t_2,u_2)$.
Thus, by induction hypothesis,
$(t_1,u_1)~(>\horco^+)\prod~(t_2,u_2)$. Hence, by (cont) and
transitivity, $t_1t_2>\horco^+ u_1u_2$.

\item By induction hypothesis, $t>\horco^+ u$. Thus, by (cont),
$\lx t>\horco^+ \lx u$.\qed
\end{enumi}
\end{proof}

We observe that, if (6) were restricted to
$(t_1>\horpo u_1\et t_2=u_2)\ou (t_1=u_1\et t_2>\horpo u_2)$, then we
would get ${>\horpo}\sle{>\horco}$, since this is the only case
requiring transitivity.

Note that $>\horco$ can be extended with the accessibility ordering
defined in Figure \ref{fig-acc-ord}. The details can be found in
\cite{blanqui06rr}.


Finally, we remark that, when restricted to first-order terms, the
recursive computability ordering is equal to the usual first-order
recursive path ordering \cite{plaisted78tr,dershowitz82tcs}, the
subterm rule being simulated by (arg) and (red).

\begin{lemma}
The relation defined in Figure \ref{fig-rco} by the rules (arg),
(decomp-symb), (call) and the rule:

\begin{center}
{\em(prec-app)}
\quad $\cfrac{f>_\cF g^{\vU\A U}\quad f\vl>\vu^\vU}{f\vl>g\vu}$
\end{center}

\noindent
is equal to the usual first-order recursive path ordering.
\end{lemma}


\section{Conclusion}

We show through various extensions how powerful is the notion of
computability closure introduced by Jean-Pierre Jouannaud and
Mitsuhiro Okada. In particular, we show how it can easily be turned
into a well-founded ordering containing Jean-Pierre Jouannaud and
Albert Rubio's higher-order recursive path ordering. This provides a
simple way to extend this ordering to richer type
disciplines. However, its definition as the closure by context of
another relation is not completely satisfactory, all the more so since
one wants to combine it with the accessibility ordering. We should
therefore try to find a new definition of HORPO that nicely integrates
the notions of computability closure and accessibility ordering in
order to capture definitions on higher-order data types (data types
with constructors having functional recursive arguments).


\begin{thebibliography}{10}

\bibitem{arts00tcs}
T.~Arts and J.~Giesl.
\newblock Termination of term rewriting using dependency pairs.
\newblock {\em Theoretical Computer Science}, 236:133--178, 2000.

\bibitem{barendregt92book}
H.~Barendregt.
\newblock Lambda calculi with types.
\newblock In S.~Abramsky, D.~Gabbay, and T.~Maibaum, editors, {\em Handbook of
  logic in computer science}, volume~2. Oxford University Press, 1992.

\bibitem{blanqui06wst-hodp}
F.~Blanqui.
\newblock Higher-order dependency pairs.
\newblock In {\em Proceedings of the 8th International Workshop on Termination,
  2006}.

\bibitem{blanqui03rta}
F.~Blanqui.
\newblock Rewriting modulo in {D}eduction modulo.
\newblock In {\em Proceedings of the 14th International Conference on Rewriting
  Techniques and Applications\em, Lecture Notes in Computer Science 2706,
  2003}.

\bibitem{blanqui00rta}
F.~Blanqui.
\newblock Termination and confluence of higher-order rewrite systems.
\newblock In {\em Proceedings of the 11th International Conference on Rewriting
  Techniques and Applications\em, Lecture Notes in Computer Science 1833,
  2000}.

\bibitem{blanqui05mscs}
F.~Blanqui.
\newblock Definitions by rewriting in the {C}alculus of {C}onstructions.
\newblock {\em Mathematical Structures in Computer Science}, 15(1):37--92,
  2005.

\bibitem{blanqui06rr}
F.~Blanqui.
\newblock {(HO)RPO} revisited.
\newblock Research Report 5972, INRIA, 2006.

\bibitem{blanqui99rta}
F.~Blanqui, J.-P. Jouannaud, and M.~Okada.
\newblock The {C}alculus of {A}lgebraic {C}onstructions.
\newblock In {\em Proceedings of the 10th International Conference on Rewriting
  Techniques and Applications\em, Lecture Notes in Computer Science 1631,
  1999}.

\bibitem{blanqui02tcs}
F.~Blanqui, J.-P. Jouannaud, and M.~Okada.
\newblock Inductive-data-type {S}ystems.
\newblock {\em Theoretical Computer Science}, 272:41--68, 2002.

\bibitem{breazu88lics}
V.~Breazu-Tannen.
\newblock Combining algebra and higher-order types.
\newblock In {\em Proceedings of the 3rd IEEE Symposium on Logic in Computer
  Science\em, 1988}.

\bibitem{breazu89icalp}
V.~Breazu-Tannen and J.~Gallier.
\newblock Polymorphic rewriting conserves algebraic strong normalization.
\newblock In {\em Proceedings of the 16th International Colloquium on Automata,
  Languages and Programming\em, Lecture Notes in Computer Science 372, 1989}.

\bibitem{breazu91tcs}
V.~Breazu-Tannen and J.~Gallier.
\newblock Polymorphic rewriting conserves algebraic strong normalization.
\newblock {\em Theoretical Computer Science}, 83(1):3--28, 1991.

\bibitem{breazu94ic}
V.~Breazu-Tannen and J.~Gallier.
\newblock Polymorphic rewriting conserves algebraic confluence.
\newblock {\em Information and Computation}, 114(1):1--29, 1994.

\bibitem{dershowitz82tcs}
N.~Dershowitz.
\newblock Orderings for term rewriting systems.
\newblock {\em Theoretical Computer Science}, 17:279--301, 1982.

\bibitem{dershowitz90book}
N.~Dershowitz and J.-P. Jouannaud.
\newblock Rewrite systems.
\newblock In J.~van Leeuwen, editor, {\em Handbook of Theoretical Computer
  Science}, volume~B, chapter~6. North-Holland, 1990.

\bibitem{dougherty91rta}
D.~Dougherty.
\newblock Adding algebraic rewriting to the untyped lambda calculus.
\newblock In {\em Proceedings of the 4th International Conference on Rewriting
  Techniques and Applications\em, Lecture Notes in Computer Science 488, 1991}.

\bibitem{dougherty92ic}
D.~Dougherty.
\newblock Adding algebraic rewriting to the untyped lambda calculus.
\newblock {\em Information and Computation}, 101(2):251--267, 1992.

\bibitem{girard71sls}
J.-Y. Girard.
\newblock Une extension de l'interpr\'etation de {G}\"odel \`a l'analyse et son
  application \`a l'\'elimination des coupures dans l'analyse et la th\'eorie
  des types.
\newblock In J.~Fenstad, editor, {\em Proc. of the 2nd Scandinavian Logic
  Symposium}, volume~63 of {\em Studies in Logic and the Foundations of
  Mathematics}. North-Holland, 1971.

\bibitem{girard72phd}
J.-Y. Girard.
\newblock {\em Interpr\'etation fonctionelle et \'elimination des coupures dans
  l'arithmetique d'ordre sup\'erieur}.
\newblock PhD thesis, Universit\'e Paris VII, France, 1972.

\bibitem{girard88book}
J.-Y. Girard, Y.~Lafont, and P.~Taylor.
\newblock {\em Proofs and Types}.
\newblock Cambridge University Press, 1988.

\bibitem{jouannaud86siam}
J.-P. Jouannaud and H.~Kirchner.
\newblock Completion of a set of rules modulo a set of equations.
\newblock {\em SIAM Journal on Computing}, 15(4):1155--1194, 1986.

\bibitem{jouannaud91lics}
J.-P. Jouannaud and M.~Okada.
\newblock Executable higher-order algebraic specification languages.
\newblock In {\em Proceedings of the 6th IEEE Symposium on Logic in Computer
  Science\em, 1991}.

\bibitem{jouannaud97tcs}
J.-P. Jouannaud and M.~Okada.
\newblock {A}bstract {D}ata {T}ype {S}ystems.
\newblock {\em Theoretical Computer Science}, 173(2):349--391, 1997.

\bibitem{jouannaud06rta-termin}
J.-P. Jouannaud and A.~Rubio.
\newblock Higher-order orderings for normal rewriting.
\newblock In {\em Proceedings of the 17th International Conference on Rewriting
  Techniques and Applications\em, Lecture Notes in Computer Science 4098,
  2006}.

\bibitem{jouannaud99lics}
J.-P. Jouannaud and A.~Rubio.
\newblock The {H}igher-{O}rder {R}ecursive {P}ath {O}rdering.
\newblock In {\em Proceedings of the 14th IEEE Symposium on Logic in Computer
  Science\em, 1999}.

\bibitem{jouannaud96rta}
J.-P. Jouannaud and A.~Rubio.
\newblock A recursive path ordering for higher-order terms in eta-long
  beta-normal form.
\newblock In {\em Proceedings of the 7th International Conference on Rewriting
  Techniques and Applications\em, Lecture Notes in Computer Science 1103,
  1996}.

\bibitem{khasidashvili90}
Z.~Khasidashvili.
\newblock Expression {R}eduction {S}ystems.
\newblock In {\em Proc. of I. Vekua Institute of Applied Mathematics},
  volume~36, 1990.

\bibitem{klop80phd}
J.~W. Klop.
\newblock {\em Combinatory Reduction Systems}.
\newblock PhD thesis, Utrecht Universiteit, The Netherlands, 1980.
\newblock Published as Mathematical Center Tract 129.

\bibitem{klop93tcs}
J.~W. Klop, V.~van Oostrom, and F.~van Raamsdonk.
\newblock Combinatory reduction systems: introduction and survey.
\newblock {\em Theoretical Computer Science}, 121:279--308, 1993.

\bibitem{kruskal60ams}
J.~B. Kruskal.
\newblock Well-quasi-ordering, the tree theorem, and vazsonyi's conjecture.
\newblock {\em Transactions of the American Mathematical Society}, 95:210--225,
  1960.

\bibitem{loria92ctrs}
C.~Loria-Saenz and J.~Steinbach.
\newblock Termination of combined (rewrite and $\lambda$-calculus) systems.
\newblock In {\em Proceedings of the 3rd International Workshop on Conditional
  and Typed Rewriting Systems\em, Lecture Notes in Computer Science 656, 1992}.

\bibitem{lysne95rta}
O.~Lysne and J.~Piris.
\newblock A termination ordering for higher order rewrite systems.
\newblock In {\em Proceedings of the 6th International Conference on Rewriting
  Techniques and Applications\em, Lecture Notes in Computer Science 914, 1995}.

\bibitem{mayr98tcs}
R.~Mayr and T.~Nipkow.
\newblock Higher-order rewrite systems and their confluence.
\newblock {\em Theoretical Computer Science}, 192(2):3--29, 1998.

\bibitem{mendler87lics}
N.~P. Mendler.
\newblock Recursive types and type constraints in second order lambda calculus.
\newblock In {\em Proceedings of the 2nd IEEE Symposium on Logic in Computer
  Science\em, 1987}.

\bibitem{mendler91pal}
N.~P. Mendler.
\newblock Inductive types and type constraints in the second-order lambda
  calculus.
\newblock {\em Annals of Pure and Applied Logic}, 51(1-2):159--172, 1991.

\bibitem{miller89elp}
D.~Miller.
\newblock A logic programming language with lambda-abstraction, function
  variables, and simple unification.
\newblock In {\em Proceedings of the International Workshop on Extensions of
  Logic Programming\em, Lecture Notes in Computer Science 475, 1989}.

\bibitem{miller91jlc}
D.~Miller.
\newblock A logic programming language with lambda-abstraction, function
  variables, and simple unification.
\newblock {\em Journal of Logic and Computation}, 1(4):497--536, 1991.

\bibitem{miller88iclp}
D.~Miller and G.~Nadathur.
\newblock An overview of $\lambda${P}rolog.
\newblock In {\em Proceedings of the 5th International Conference on Logic
  Programming\em, MIT Press, 1988}.

\bibitem{nipkow91lics}
T.~Nipkow.
\newblock Higher-order critical pairs.
\newblock In {\em Proceedings of the 6th IEEE Symposium on Logic in Computer
  Science\em, 1991}.

\bibitem{okada89issac}
M.~Okada.
\newblock Strong normalizability for the combined system of the typed lambda
  calculus and an arbitrary convergent term rewrite system.
\newblock In {\em Proceedings of the 1989 International Symposium on Symbolic
  and Algebraic Computation\em, ACM Press}.

\bibitem{plaisted78tr}
D.~A. Plaisted.
\newblock A recursively defined ordering for proving termination of term
  rewriting systems.
\newblock Technical report, University of Illinois, Urbana-Champaign, United
  States, 1978.

\bibitem{sakai01rpc}
M.~Sakai and K.~Kusakari.
\newblock On new dependency pair method for proving termination of higher-order
  rewrite systems.
\newblock In {\em Proceedings of the 1st International Workshop on Rewriting in
  Proof and Computation, 2001}.

\bibitem{tait67jsl}
W.~W. Tait.
\newblock Intensional interpretations of functionals of finite type {I}.
\newblock {\em Journal of Symbolic Logic}, 32(2):198--212, 1967.

\bibitem{tait72lc}
W.~W. Tait.
\newblock A realizability interpretation of the theory of species.
\newblock In R.~Parikh, editor, {\em Proceedings of the 1972 Logic Colloquium},
  volume 453 of {\em Lecture Notes in Mathematics}, 1975.

\bibitem{vandepol93hoa}
J.~van~de Pol.
\newblock Termination proofs for higher-order rewrite systems.
\newblock In {\em Proceedings of the 1st International Workshop on Higher-Order
  Algebra, Logic and Term Rewriting\em, Lecture Notes in Computer Science 816,
  1993}.

\bibitem{oostrom95hoa}
V.~van Oostrom.
\newblock Development closed critical pairs.
\newblock In {\em Proceedings of the 2nd International Workshop on Higher-Order
  Algebra, Logic and Term Rewriting\em, Lecture Notes in Computer Science 1074,
  1995}.

\bibitem{oostrom94phd}
V.~van Oostrom.
\newblock {\em Confluence for Abstract and Higher-Order Rewriting}.
\newblock PhD thesis, Vrije Universiteit Amsterdam, The Netherlands, 1994.

\bibitem{peterson81jacm}
G.~Peterson and M.~Stickel.
\newblock Complete sets of reductions for some equational theories.
\newblock {\em Journal of the ACM}, 28(2):233--264, 1981.

\bibitem{raamsdonk96phd}
F.~van Raamsdonk.
\newblock {\em Confluence and Normalization for Higher-Order Rewriting}.
\newblock PhD thesis, Vrije University Amsterdam, The Netherlands, 1996.

\bibitem{walukiewicz03phd}
D.~Walukiewicz-Chrz\k{a}szcz.
\newblock {\em Termination of Rewriting in the {C}alculus of {C}onstructions}.
\newblock PhD thesis, Warsaw University, Poland and Universit\'e d'Orsay,
  France, 2003.

\bibitem{walukiewicz03jfp}
D.~Walukiewicz-Chrz\k{a}szcz.
\newblock Termination of rewriting in the {C}alculus of {C}onstructions.
\newblock {\em Journal of Functional Programming}, 13(2):339--414, 2003.

\end{thebibliography}

\end{document}